\documentclass[showpacs,preprintnumbers,aps]{revtex4}
\usepackage{amsfonts}
\usepackage{amssymb}
\usepackage{amsmath}
\usepackage[all]{xy}
\begin{document}
\preprint{USM-TH-178}
\newcommand{\ba}{\begin{eqnarray}}
\newcommand{\ea}{\end{eqnarray}}
\newcommand{\be}{\begin{equation}}
\newcommand{\ee}{\end{equation}}
\newcommand{\bib}{\bibitem}
\newcommand{\ed}{\end{document}}
\newcommand{\nn}{\nonumber\\}
\newcommand{\fr}{\frac}
\newcommand{\wt}{\widetilde}

\title{Physical and mathematical evidences for a negative-rank tensor}
\author{Patricio Gaete}
\email {patricio.gaete@usm.cl} \affiliation{Departamento de
F\'{\i}sica, Universidad T\'ecnica F. Santa Mar\'{\i}a,
Valpara\'{\i}so, Chile}
\author{Clovis Wotzasek}
\email {clovis@if.ufrj.br}
\affiliation{Instituto de F\'\i sica, Universidade Federal do Rio de
Janeiro, Brazil}
\date{\today}
\begin{abstract}
We propose and study the properties of a new potential demanded by the
self-consistency of the duality scheme in electromagnetic-like field
theories of totally anti-symmetric tensors in diverse dimensions.
Physical implications of this new potential is manifest under the
presence of scalar condensates in the Julia-Toulouse mechanism for
the nucleation of topological defects with consequences for the
confinement phenomenon.
\end{abstract}
\pacs{11.10.Ef, 11.10.Kk}
\maketitle

Physics is an experimental science and theoretical advances are
generally driven by careful and delicate observations coming from
lab measurements aimed to gain further insights in the
overall structure of the theories which describes physical
phenomena. Most of the time experimental and theoretical efforts
feed each other back to advance in the general goal of physics - the
description and interpretation of Nature. There are, however,
occasions when adjusts in the theoretical framework become necessary
to preserve the structure of the basic postulates under new
observations. Sometimes these purely theoretical works may come to
deeply modify the way we understand the natural phenomena. Probably
the most striking examples are (i) Einstein's Special Relativity,
inspired by the existence of a dramatic coincidence in the Faraday's
law \cite{AE} and (ii) Dirac's charge-monopole quantization \cite{PAMD}
suggested by an asymmetry in Maxwell's equations, opening the
field of topology to theoretical physics. In this Letter we want to
report on a lack of consistency of the duality symmetry in the
generalized electromagnetic framework of theories described by
massless anti-symmetric tensors of arbitrary ranks in diverse
dimensions \cite{reviews}. We have found that a new and unexpected object becomes
necessary in order to maintain the tensorial structure of the
duality symmetry. It is important to mention that our present study
is not just a theoretical curiosity. We have in mind physical
implications for the condensation of topological defects and duality to various field theories
in which the notion of confinement is of fundamental
phenomenological and experimental significance \cite{euro}.

The confinement of the fundamental constituents of matter is
probably one of the fundamental and long-standing issues of theoretical
physics whose solution has evaded complete comprehension despite intense
theoretical and experimental effort. In particular, field theories that yield a linear
potential are very important to particle physics, since those theories
may be used to describe the confinement of quarks and gluons and be
considered as effective theories of quantum chromodynamics.
In the past, many authors have shown that the condensation of topological defects such as vortices and monopoles may lead to quark confinement \cite{quark}. Due to
intense interest mainly in string related topics, these studies have been
extended out of the four dimensional domain to theories of antisymmetric
tensors of arbitrary ranks in arbitrary space-time dimensions that appear as
low-energy effective field theories of strings. This has also helped us to
gain insights over the mechanisms of confinement in different contexts  \cite{ast}.

Duality, on the other hand, is a concept of paramount and ever-increasing importance
playing a fundamental role in distinct areas of nowadays physics \cite{duality}.
The electromagnetic duality is probably the main paradigm of this
notion displaying a beautiful symmetry between the electric and
magnetic sectors of the Maxwell's equations \cite{GR}.  Because of the major
role played by the extended electromagnetic theories of
anti-symmetric tensors in QCD, Super-Gravity and String Theory,
duality has been intensely examined in those scenarios in recent
years, including couplings with extended sources (branes) of proper
dimensions.

Certainly the best way to describe tensors in this extended
electromagnetism is by use of differential forms.  However, in order
to allow access to a larger audience we shall avoid the use of such
a language or, in occasions, use it in a more relaxed and intuitive
form \footnote{We shall follow the simplified notation introduced in \cite{euro} which basically do not distinguish between a form and its components.}.
Let us consider then theories described by a rank-$p$ totally
anti-symmetric potential $A_p=A_{\mu_1\cdots\mu_p}$ whose
field-strength reads, as usual
\be \label{10} F_{p+1}(A_p) \equiv
\partial_{[\mu_1} A_{\mu_2\cdots\mu_{p+1}]}\, ,
\ee
being proportional to the exterior derivative of $A_p$, with
Greek indices taking values over the space-time dimensions $\mu =
0,1,\cdots,D-1$ \footnote{In differential form language the field
tensor is given by the exterior derivative of the potential as
$F_{p+1}= d\, A_p$.}. The action controlling the classical dynamics
is given by the standard textbook form
\be \label{20} {\cal S}_p = {\cal N}_p
\int F_{p+1}F^{p+1}\, , \ee
where ${\cal N}_p$ is a proper normalization constant. This is also the starting point to the canonical quantization program.
It is noticeable that in the free action the potential tensor $A_p$ itself
is not present, only its exterior derivative.

For $p=0$, the potential $A_0$ is just a scalar field whose field
equations are described by the massless Klein-Gordon operator. The
$p=1$ case describes the usual Maxwell case with the vector
potential $A_1\equiv A_\mu$ giving origin to the electric/magnetic
fields described by the rank-two field tensor $F_2(A_1)\equiv
F_{\mu\nu}(A_\mu)$ as, \be \label{30} F_{\mu\nu}=\partial_\mu A_\nu
- \partial_\nu A_\mu \, , \ee whose action and field equations are
invariant under the usual gauge symmetry, \be \label{40} \delta
A_\mu = \partial_{\mu} \xi \, . \ee Another interesting example is
the $p=2$ case described by the Kalb-Ramond field $A_2 \equiv
A_{\mu\nu}$ whose action is invariant under the rank-two gauge
transformation $\delta A_{\mu\nu} =
\partial_{[\mu}\, \xi_{\nu ]}$. In general, the rank-$p$ theory is
invariant under an extended gauge transformation as, \be
\label{50} \delta A_p = \partial_{[\alpha}\,\xi_{p-1 ]} \equiv F_p
\left(\xi_{p-1}\right)\, , \ee involving a gauge function
$\xi_{p-1}$ of one degree lesser.

A canonical analysis, taking into account the gauge constraints of
these theories reveals that the model described by the tensor
potential $A_p$ possess $C_p^{D-2}$ degrees of freedom, where \be
\label{60} C_p^{D-2} = \frac {(D-2)!}{ p\,! \, (D-p-2)!}\, , \ee is
the binomial coefficient. For vector potentials in the Maxwell case,
the number of degrees of freedom is just $(D-2)$ giving the two well
known polarization of the four-dimensional photon, just a single
degree of freedom in the tree-dimensional case and no transverse
propagating degrees of freedom for a theory defined on the line, as it should be
expected. The Pascal triangle symmetry rule demands that \be
\label{70} C_p^{D-2} = C_q^{D-2}\, , \ee which is valid if \be
\label{80} p+q+2 = D\, . \ee There should exist therefore, a dual tensor potential $\tilde A_q$,
to be properly defined below, that although being in general of distinct rank,
will propagate exactly the same number of transverse degrees of
freedom as the original $A_p$ potential which suggests that they are
indeed related by duality. In fact, one can show through the duality
mapping, a Legendre transformation followed by the solution of a
simple constraint, that this assumption is indeed correct. We shall
refer to the simple rule (\ref{80}) as the {\it massless duality condition}.
An important instance is the $p=1$ case where the dual tensor is
also a vector for $D=4$ but it is a scalar for $D=3$. It is
fundamental to notice that apparently the vector Maxwell theory on the line
has no dual-partner. This difficulty is generic for the whole class
of extended electromagnetism and it is the main point of discussion
in this paper.

Let us next consider the inclusion of sources, the extended objects
known as branes. They may appear as topological defects coming from
the low-energy sector of theories having broken symmetries, where
the defects are solitonic solutions. Current conservation demands
the branes to be closed.  Closed $(p-1)$-branes are characterized by
their charge, say $e$, and the Chern-Kernel $\Lambda_{q+1}$ \cite{HL}, giving
the localization of the brane, such that their (conserved)
$p$-current may be written as \footnote{According to the
simplified notation introduced in \cite{euro}, we have
$J^{\mu_1\cdots\mu_p} =
\epsilon^{\mu_1\cdots\mu_p\alpha\nu_1\cdots\nu_{q+1}}
\partial_\alpha \Lambda_{\nu_1\cdots\nu_{q+1}}$.}, \be \label{90}
J^p(\Lambda) = \epsilon^D \partial_1 \Lambda_{q+1}\, . \ee This
form automatically displays the closeness of the source and shows
an invariance under \be \label{100} \delta \Lambda_{q+1} =
\partial_{[1}\,\chi_{q]}\, , \ee meaning that its $p$-dimensional
Dirac-brane, whose Chern-Kernel is $\chi_q$ is physically
unobservable. The ``smallest" brane is the {\it instanton} whose
dimension is $(-1)$.  The usual point-charge is a zero-brane and a
string is an one-dimensional brane. The ``biggest" brane fitting
in a $D$-dimensional space-time is the $(D-2)$-brane whose
Chern-Kernel is $\Lambda_0$ and its $d$-current is $J^d = \epsilon^{d,1}\partial_1 \Lambda_0$.

A $(p-1)$-brane is an extended charge which, upon time-evolution,
describes a $p$-dimensional object, its world-volume current. These
objects may therefore couple transversally to a rank-$p$ potential
and longitudinally to a rank-$q$ potential, with $p$ and $q$ keeping
a relation as in (\ref{80}).  In the usual electromagnetic {\it
jargon}, the $A_p$ potential ``sees" the $(p-1)$-brane as an
electric brane while its dual tensor, $\tilde A_q$ experiences it as
magnetic. Therefore, the potential $A_p$ may couple electrically and
magnetically to two distinct branes of dimensions $(p-1)$ and
$(q-1)$, respectively, according to the following diagram,
 \be
\label{diagrama1}
\xymatrix{*+[F]{{(q-1)\,;\, g \, ;\,\Omega_{p+1}}}  \ar[r]& A_p
\ar[l]^-{\rm{MC}} \ar[r]_-{\rm{EC}} & *+[F]{{(p-1)\, ;\,e\, ;\,
\Lambda_{q+1}}}  \ar[l]
          }\,\, .
\ee
Here, the $e$-charged $(p-1)$-brane is described by its Chern-Kernel
$\Lambda_{q+1}$ while the $g$-charged $(q-1)$-brane is described by
$\Omega_{p+1}$.
The classical action for this system is \cite{PAMD,GR}
\be
\label{120}
{\cal S} = \int \left\{\left[F_{p+1}(A_p) - g\, \Omega_{p+1}\right]^2
 + e A_p \,\epsilon^D \partial_1\Lambda_{q+1}\right\}\, ,
\ee
whose equations of motion read
\ba
\label{130}
\partial_1 \, F^{p+1} &=& e\, J^p(\Lambda)\nonumber\\
\partial_1 \, \mbox{}^*F^{q+1} &=& g\, J^q(\Omega)\, ,
\ea where the Hodge dual tensor is defined as
\be \label{140}
\mbox{}^* F^{q+1} = \epsilon^{q+1,p+1} F_{p+1}\, . \ee
These equations
display a pair of gauge symmetries \cite{ezawa,kleinert}. The usual (electric) gauge
transformation (\ref{50}) leaves invariant both terms in
(\ref{120}) thanks to the conservation of the electric current (or
the closeness of the brane). There is also a ``magnetic" like
gauge transformation combining $\delta \Omega_{p+1} =
\partial_{[1}\,\tilde\chi_{q]}$
with $\delta A_p = g \tilde\chi_p$ which leaves invariant the
first term in (\ref{120}). The invariance of the second term under
the magnetic gauge transformation is only obtained if the charges
of the branes are quantized according to Dirac's condition \cite{diraccondition},
\be
\label{150} e\, g = 2\pi\, n\;\; ;\;\; n\in {\cal Z}\, . \ee

Duality maps the $A_p$ potential into its dual $\tilde A_q$ and
reverts the couplings, as shown in the dual action \be \label{160}
\tilde{\cal S} =  \int \left\{\left[\tilde F_{q+1}(\tilde A_q) -
e\, \Lambda_{q+1}\right]^2 + g \tilde A_p \epsilon^D
\partial_1\Omega_{p+1}\right\}\, , \ee and, thanks to Dirac's
quantization rule (\ref{150}), an interacting brane-brane term is
not induced into the dual action.
In a certain sense, the duality map and the charge quantization
constraint displayed by these electromagnetic-like systems are
restricted by the condition (\ref{80}). The massless duality rule
(\ref{80}) contains indeed quite interesting information. It
shows, for instance, that for $D=4$ the usual vector potential is dual to another
vector potential, a well known fact. Since both dual tensors are
of the same rank we call this situation as self-duality. However,
even in four-dimensions, there is another, much less known,
possibility for duality, between a scalar field and a rank-2
Kalb-Ramond potential \cite{duality2}.

Now we make an important remark. Notice that for every dimension
there is a highest-rank potential, $A_{D-1}$, whose field strength
must be proportional to the totally anti-symmetric Levi-Civita
symbol, \be\label{170} F_D = f \, \epsilon_D \ee These fields
carry no propagating degrees of freedom but they seem to be of
importance for the gauge representation of the cosmological
constant with important consequences. For instance, in \cite{AAS}
the rank-3 tensor of a four-dimensional space-time was suggested as a possible candidate for dark
matter or dark energy. This maximum rank potential couples
transversally (electrically) to the maximum-dimension brane.
However, there is no  longitudinal coupling for this maximum rank potential because for that it
would be necessary a brane of dimension $(-2)$, see Eq.(\ref{diagrama1}).

What seems peculiar here is that according to the massless duality
rule (\ref{80}), the tensor dual to the maximum-rank potential
$A_{D-1}$ should have negative rank. In fact it should be a
$(-1)$-rank potential, say $A_{-1}$. If such an object is expected to exist, as demanded by duality,
it would couple longitudinally (magnetically) with the
maximum-dimension brane but the transverse coupling (electric)
would demand a $(-2)$-brane. The existence of a negative rank
tensor (NRT) would therefore complete the duality scheme in
electromagnetic-like theories. This situation is illustrated in
the diagram below for the special case of $D=4$, where the maximum-dimension brane is the membrane and the maximum rank potential is the 3-tensor $A_3$,
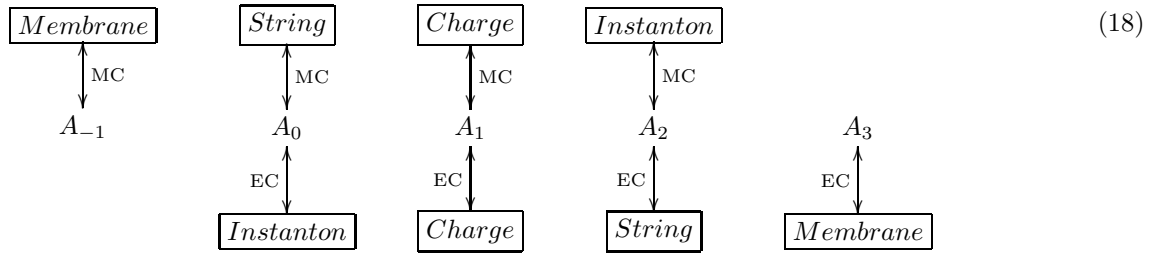
\be
\label{diagrama2} \xymatrix{ *+[F]{Membrane}\ar[d]^-{\rm{MC}} &
*+[F]{String}\ar[d]^-{\rm{MC}} & *+[F]{Charge}\ar[d]^-{\rm{MC}} &
*+[F]{Instanton}\ar[d]^-{\rm{MC}}&
\\
A_{-1} \ar[u]
& A_0 \ar[u]\ar[d] &  A_1 \ar[u]\ar[d] &  A_{2} \ar[u]\ar[d] & A_{3}
\ar[d] \\
& *+[F]{Instanton}\ar[u]^-{\rm{EC}} & *+[F]{Charge}\ar[u]^-{\rm{EC}}
 & *+[F]{String}\ar[u]^-{\rm{EC}} & *+[F]{Membrane}\ar[u]^-{\rm{EC}}}
\ee

The important point to be observed at this juncture is that
according to standard classification of differential geometry the
scalar field, say $A_0$, is the one expected to be the minimum-rank tensor so that there is no room for the negative rank tensor \cite{nakahara}.
So, at this point we have to face the dilemma of either break the
duality scheme (\ref{80}), keeping it asymmetrical, or take a rather
pragmatic point of view and extend the well established differential
geometry classification by introducing a new and unusual object. We
decided to take chance with the latter case.

In fact, the differential in favor of the NRT seems to be the
absence of a transverse coupling for it. The existence of such a coupling would demand the
presence of the tensor itself while the longitudinal coupling only
demands the presence of its (exterior) derivative and not the tensor itself. This fact
suggests that adopting as the definition for the NRT just that its
exterior derivative leads to a zero-rank tensor, \be \label{180} A_0
= d\, A_{-1} \ee suffices to obtain its proper action both free or
longitudinally coupled to the maximum-dimension brane. The NRT {\it
per se} is never needed. Even when coupled to the maximum-brane its action would only contain the scalar field
without any derivative. Therefore it will not propagate any degree
of freedom, as expected from duality arguments. Incidentally, it is interesting to observe that such definitions are not new
in Physics. Let us recall the example of the Dirac ``delta function"
which is only properly defined ``inside an integral operator".
Similarly, the NRT as well is only meaningfully defined ``upon
application of the exterior derivative operator."

Next, let us discuss a more physically motivated aspect of the
subject. To this end we consider the nucleation of topological
defects in the phenomenological framework developed by Quevedo and
Trugenberger (QT) that considers condensation of topological
defects induced by quantum fluctuations \cite{QT}.  Upon
nucleation, the condensate of the g-charged $(q-1)$-dimensional
branes will absorb the degrees of freedom of its longitudinally
coupled tensor $A_p$, see Eq.(\ref{120}).  The resulting condensate becomes massive
while its Chern-Kernel is elevated to the status of propagating
field describing the long-wavelength hydrodynamical fluctuations
of the condensate. This phenomenon has been christened as
Julia-Toulouse mechanism (JTM) \cite{JT} in (QT).  Its main
characteristic, that distinguishes it from its dual, the Higgs
mechanism, is the occurrence of rank-jump - the massive condensate
has the rank given by the Chern-Kernel which is one degree higher
than the original electromagnetic potential while in the Higgs
mechanism the massive condensate inherits the degree of the
tensor.

The JTM has drastic and important consequences for the electric
objects living inside the condensate in the form of the
confinement phenomenon known as dual superconductivity. To see
this consider the g-charged brane which, before condensation, was
coupled transversally to a rank-$q$ tensor, $B_q$ and
longitudinally to a tensor $A_p$, whose action reads, \be
\label{190} {\cal S} = \int \left\{\left[F_{p+1}(A_p) - g\,
\Lambda_{p+1}\right]^2 + g B_q \,\epsilon^D
\partial_1\Lambda_{p+1} + F_{q+1}^2(B_q)\right\}\, . \ee After
condensation, the action involving the condensate
and the $B_q$ tensor should read \cite{confinement} \be\label{200} {\cal S}_{cond} =
\int \left[F_{p+2}^2(\Lambda_{p+1}) + m^2 \, \Lambda_{p+1}^2  + g
B_q \,\epsilon^D \partial_1\Lambda_{p+1}+ F_{q+1}^2(B_q)\right] \ee
displaying two phenomenological parameters $m$ and $g$ with mass
dimension and a $B\wedge F(\Lambda)$ coupling, with important
consequences. To see that after condensation electrically charged
objects living inside the condensate get irremediably confined we
compute the effective theory for the $B_q$ tensor by integrating out
the condensate degrees of freedom $\Lambda_{p+1}$, to obtain
\cite{confinement}, \be\label{210} {\cal L} = \frac 12 F_{q+1}(B_q)\left[1 +
\frac{g^2}{\Box + m^2}\right] F_{q+1}(B_q)\, . \ee This allows us to
illustrate the discussion by computing the interaction energy
between static $q$-charged point-like sources a distance $L$ apart, for the
theory under consideration. As is well known, the interaction
between two static charges is generally investigated in terms of
Wilson loops. We can, however, more directly recover the interaction
potential by using the gauge-invariant variables formalism along the
lines of Ref.\cite{confinement}. In such a case, the corresponding static
potential is given by \be\label{220} V(L) = \cdots + \sigma(m) L\,
,\ee where the ``string" tension $\sigma(m,q)$ is expected to depend
both on the mass $m$ of the condensate and on the charge of the probes. The ellipsis
represent the screening part that is very much
dimensional dependent and must be computed case by case. An specific example will be given below.
This confinement property is independent of
the degree of the condensate that, in its turn, only depends of the
dimension of the condensing brane. This simple explanation for the
confinement phenomenon as the result of quantum fluctuations impart
important physical significance to the QT phenomenology of the JTM.
In \cite{confinement} the static potential above has been proposed as a way to measure the phenomenological parameter $m$, the mass of the condensate, in the theory put forward in \cite{QT}.

One can expect therefore to obtain condensates of distinct degrees
for each spacetime dimension. The degree of the highest condensate is $(D-1)$ coming from the condensation of magnetic instantons. Clearly a $D$-degree condensate cannot be induced since that requires a brane with dimension (-2). An interesting instance in
$D=4$ is given by the condensation of a magnetic string, described
by a degree-one Chern-Kernel that couples longitudinally with a
scalar field and transversally with a Kalb-Ramond tensor. After
the nucleation, the Chern-Kernel of the magnetic string becomes a
vector field that, according to the JTM, absorbs the single degree
of freedom of the scalar to become a vectorial massive condensate.

What is however quite peculiar in this case is the fact that,
according to the QT scheme, the condensate of lowest degree has
vectorial character.  There is no room for a scalar condensate
unless the NRT is included, as demanded by the duality scheme.
Indeed the existence of a scalar condensate would necessarily come
from the condensation of the biggest brane, a $(D-2)$-dimensional
object, whose zero-degree Chern-Kernel would then lead to the
proper condensate. The presence of a NRT would allow for the application of the JTM, as proposed by QT, to the biggest brane. The existence of such a scalar condensate is demanded by consistency of the duality mapping. By this we mean that the condensation of this very same brane when examined from the dual point of view, the corresponding Higgs mechanism, is indeed possible.

A well known instance of a scalar condensate interacting with
electric charges is given by investigations in $D=2$, the long-time
used laboratory for quantum field theory, known as {\it massive
Schwinger model} \cite{SC1}.  Let us examine how the phenomenology of this
model is reobtained from the QT-JTM point of view. An
electromagnetic-like field theory can be defined on the line with
three tensors ($A_{-1}$, $A_0$, $A_1$) and has only enough ``room"
for instantons and charges, higher dimensional branes not being
allowed. By considering the above general scheme for condensation of
magnetic monopoles in $D=2$, we obtain the following JT action, \be
\label{230} {\cal S} = \int \left[\left(F_{0}(A_{-1}) - g\,
\Lambda_{0}\right)^2 + g B_1 \,\epsilon^D \partial_1\Lambda_{0}+
F_{2}^2(B_1)\right] \stackrel{JTM}{\rightarrow}\int
\left[F_{1}^2(\Lambda_{0}) + m^2 \, \Lambda_{0}^2  + g B_1
\,\epsilon^D \partial_1\Lambda_{0} + F_{2}^2(B_1)\right] \ee which
we recognize as the action for the massive Schwinger model
\footnote{There is however a subtlety here. The QT approach is just
a phenomenological way to incorporate quantum fluctuations of
topological defects into the theory. Because of the strong demands
they made on gauge invariance, Lorentz symmetry and linearity only
part of this contribution is obtained, although with enough physical
content to produce confinement. For the case at hand the complete
contribution would give the sine-Gordon potential whose first
non-trivial contribution gives the mass term.}. In the QT/JTM
scheme, the (bosonized) scalar field which is induced by quantum
fluctuations of the magnetic monopoles, describes the degrees of
freedom of the condensate. The effective theory for electric charges
exchanging (vectorial) photons inside such a condensate is given by
Eqs.(\ref{190}-\ref{210}) with $p= -q = -1$. The static potential is
obtained studying, \be\label{240} {\cal L} = \frac 12
F_{2}(B_1)\left[1 + \frac{g^2}{\Box + m^2}\right] F_{2}(B_1) + B_1\,
J^1 \ee where $J_1$ describes two static heavy point charges. As in
the previous example, the interaction energy between pointlike
sources clearly displays the confinement phenomenon \cite{GaeteSch}:
\be V(L)=\frac{{q^2 }}{{2\lambda }}\left( {1 +
\frac{{m^2 }}{{\lambda ^2 }}} \right)\left( {1 - e^{ - \lambda L} }
\right) + \frac{{q^2 }}{2}\left( {1 - \frac{{g^2 }}{{\lambda ^2 }}}
\right)L \, ,\ee with $\lambda ^2  = g^2  + m^2$ which displays both the screening and the
confining part of this interaction. Of course, if we let the
system returns to the dilute phase, i.e., $m\to 0$, the static
potential above shows that confinement disappears, as expected.

It is interesting, at this juncture, to make connection between
the monopole condensation described above involving the NRT and
the the fermion condensate described by bosonization of the
massive Thirring model (MTM) \cite{SC2}. In fact one can understand a mass
term in the MTM model in terms of perturbed Conformal Field
Theory. Recall that a cosine potential in the sine-Gordon model is
a perturbation from a massless free scalar theory. A mass term in
the action of the MTM model corresponds to a cosine potential
through the abelian bosonization, and we can identify a mass term
in the MTM model as a perturbation from a massless Thirring model.
Correspondingly, the monopole effect realized by a cosine
potential in the sine-Gordon model is described by a mass term in
the MTM model. Such a correspondence leads us to expect the
relation between the monopole condensate in compact QED and the
fermion condensate in the MTM model \cite{YS}.

In summary, we have shown that consistency of the duality picture
in the generalized electromagnetism of totally anti-symmetric
tensors in arbitrary dimensions requires the existence of a new and
unusual potential, the negative-rank tensor. In this work we have shown that this object is allowed
in the duality scheme because only its (exterior) derivative is
required on the action and field equations, not the tensor itself. This feature allows us to incorporate this object into
the well defined scheme of differential geometry without
destroying the well established classification scheme: it suffices
that the exterior derivative of the NRT maps it into a zero-form.
A physical motivation for the existence of this object is given in
the context of the Julia-Toulouse mechanism that describes the
condensation of topological defects. It was observed that without
the presence of the NRT it is unable to generate condensates of
scalar nature, therefore breaking duality with the Higgs
mechanism. Theories that present confinement due to existence of a
scalar condensate, such as the $D=2$ sine-Gordon model, would not
otherwise fit into the JT scheme without the presence of such a
tensor.

\section{ACKNOWLEDGMENTS}

 P. G. was partially supported by FONDECYT (Chile)  grants  1050546 and 7050201. C. W. thanks the Physics Department of the Universidad T\'ecnica F. Santa Mar\'{\i}a for hospitality, he also thanks CNPq and PRONEX for financial support.

\end{document}